# Quantum-limit phenomena and bandstructure in the magnetic topological semimetal EuZn$_2$As$_2$

Joanna Blawat[1,2], Smita Speer[2], John Singleton[3], Weiwei Xie[4], Rongying Jin[1,2*]

[1]Center for Experimental Nanoscale Physics, Department of Physics & Astronomy, University of South Carolina, Columbia, SC 29208, USA

[2]Department of Physics & Astronomy, Louisiana State University, Baton Rouge, LA 70803, USA

[3]National High Magnetic Field Laboratory, Los Alamos National Laboratory, Los Alamos, New Mexico 87545, USA

[4]Department of Chemistry, Michigan State University, East Lansing, MI 48824, USA

*Corresponding author: rjin@mailbox.sc.edu

## Abstract

We have experimentally investigated the low-temperature (0.6 K) electronic and magnetic properties of the layered antiferromagnet EuZn$_2$As$_2$ in pulsed magnetic fields of up to 60 T at a temperature of 0.6 K, giant positive magnetoresistance (MR) is observed above $\mu_0 H \approx 20$ T, a regime in which the spins are already fully polarized. Both magnetic torque and proximity detector oscillator (PDO) data show no corresponding anomaly at or close to this field. By analyzing the quantum oscillations observed in the MR and PDO frequency, we find that (1) the oscillation frequency $F = 46 \pm 6$ T for $H // c$ and $42 \pm 2$ T for $H // ab$; (2) the corresponding Berry phase is close to $\pi$ for $H // c$, implying a nontrivial topology; and (3) the large linear MR at high fields corresponds to the quantum limit (*i.e.,* only the last Landau level being occupied). From these observations we conclude that the linear MR can be understood by considering diffusing cyclotron centers in the quantum limit. Our findings help understand the intimate relationship between magnetism and electronic topology in EuZn$_2$As$_2$ under extremely high fields and suggest reasons for the emergent behavior in the quantum limit.

**Introduction**

Understanding the interplay between magnetism and non-trivial electronic topology is a new frontier in condensed matter physics [1–3]. Extensive research has focused on magnetic topological semimetals (MTSMs) with linear electronic energy dispersion in momentum space [4,5]. Such materials offer unique opportunities to manipulate the electronic band structure and its topology by changing the spin configuration, and are therefore a potential platform for designing new spintronic devices [6–8].

Crucial in understanding the formation of non-trivial topological states in semimetals are spin-orbit coupling (SOC), crystal symmetry ($\mathcal{P}$), and magnetic ordering which breaks time-reversal symmetry ($\mathcal{T}$). The Dirac state is protected by both $\mathcal{P}$ and $\mathcal{T}$, and can be transformed to a Weyl state through breaking either $\mathcal{P}$ or $\mathcal{T}$ [12]. The linear band dispersion in Dirac and Weyl semimetals may contribute to many unusual transport properties, such as large linear magnetoresistance (MR) [9], ultrahigh mobility [10–13], chiral anomalies [14,15], and the anomalous Hall effect [16,17]. However, the observation of new quantum phenomena in MTSMs often requires extreme conditions such as high magnetic fields and low temperatures. In addition to quantizing the quasiparticle energy into Landau levels (LLs), external fields may change the magnetic structure of MTSMs, resulting in new states with unusual properties. For example, the quantum Hall effect is only observed in the canted antiferromagnetic state accessed by applying magnetic field in $EuMnBi_2$ [18]. On the other hand, higher magnetic fields increase the LL degeneracy, eventually resulting in only the lowest Landau level (LLL) being populated. The combination of the relativistic nature of Weyl/Dirac fermions with these quantum-limit conditions has been linked to emergent quantum phenomena such as interlayer quantum tunneling transport in $YbMnBi_2$ [19], a magnetic torque anomaly in the Weyl semimetal NbAs [20], Weyl node annihilation in noncentrosymmetric TaP [21], and linear MR after reaching the LLL in $Cd_3As_2$ [22].

Among known MTSM candidates, Eu-based compounds are unique, as the orientation of the Eu moments is sensitive to external magnetic field; this scenario is ideal for investigating the effect of magnetic symmetry [18,23,24]. Of particular interest is the $EuM_2As_2$ (M = Cd, In, Zn *etc.*) material family with nonsymmorphic time-reversal symmetry [23–25]. A Dirac state is predicted in the antiferromagnetically (AFM) ordered phase of $EuCd_2As_2$ ($T_N \approx 8.5$ K) [24]. Experimentally, Weyl nodes have been observed above $T_N$ due to strong ferromagnetic (FM) fluctuations [26] or the application of magnetic field [24]. For $EuZn_2As_2$, both $T_N$ and the FM fluctuations regions are doubled compared to EuCd$_2$As$_2$ [23], making it a better platform for studying the interplay between magnetism and topology. In this article, we report the electronic and magnetic properties of EuZn$_2$As$_2$ under pulsed magnetic fields of up to 60 T. Giant MR is observed above $\mu_0 H \approx 20$ T. By analyzing Shubnikov-de Haas (SdH) and de Haas-van Alphen (dHvA) oscillations, the 1$^{st}$ LL is reached at $\approx 50$ T for $H \parallel c$, above which the MR becomes linear. Our findings help understand the interplay between magnetism and topological properties and emergent behavior beyond the quantum limit.

**Methods**

Single crystals of EuZn$_2$As$_2$ were synthesized by the flux method as described in Ref. [23]. The low-field magnetic properties were measured in a *Quantum Design* Magnetic Properties Measurement System (MPMS – 7 T), while the high-field data were taken at the pulsed-field facility of the National High Magnetic Field Laboratory (NHMFL, Los Alamos). Standard four-probe techniques were used to measure the MR. Four thin Pt wires were placed on a sample using silver epoxy, and each Pt wire was attached to 50-gauge copper wire. The copper wires were twisted in pairs (current pair and voltage pair) to reduce electrical noise. Samples were placed on a cryogenic goniometer (fabricated using additive manufacturing techniques [27]) for measurements at different angles in the pulsed magnetic fields. An AC current of 0.23 mA was applied at 50 kHz frequency; care was taken to ensure that this did not cause heating in the sample. The torque measurements were carried out in magnetic fields of up to 60 T by mounting a 50 μm long crystal on a piezoresistive cantilever [28]. For performing proximity detector oscillation (PDO) measurements, a sample was placed on the top of an 8-turn coil made of 46-gauge copper wire. The coil is connected to the PDO circuit which has a resonant frequency in the range of 22 -30 MHz determined by the sample's conductivity and, to a lesser extent, its magnetic susceptibility [28].

**Results and Discussion**

EuZn$_2$As$_2$ has a trigonal crystal structure (P-*3m1*, #164) and orders in an A-type spin structure at $T_N = 19$ K [23]. Figure 1(a) presents the angle dependence of the inverse magnetic susceptibility ($\chi^{-1}$) measured at $\mu_0 H = 0.1$ T (see inset for the definition of the angle ϕ). Note that $\chi^{-1}$ decreases linearly with decreasing temperature above $\approx 200$ K, which can be fitted by the Curie-Weiss formula $\chi(T) = \chi_0 + C/(T - \theta)$ ($\theta$ is the Curie-Weiss temperature, $C = \frac{\mu_{eff}^2 N_A}{3k_B}$ is the Curie constant, $\mu_{eff}$ is the effective moment, $N_A$ is the Avogadro constant, and $k_B$ is the Boltzmann constant). The obtained $\theta$ and $\mu_{eff}$ are plotted as a function of angle $\phi$ in the inset of Figure 1(a). The positive $\theta$ suggests ferromagnetic interactions between Eu ions, with the largest value at $\phi = 45°$. Conversely, $\mu_{eff}$ exhibits the smallest value at $\phi = 45°$, likely due to the variation of the spin-orbit coupling (SOC). Well below $T_N$ [23], $\mu_{eff}$ obtained at 2 K shows a similar angle dependence. Figure 1(b) presents the field dependence of the magnetization at $\phi = 45°$, which initially increases linearly with field then becomes saturated ($\mu_{sat}$) at $\mu_0 H_c = 2.4$ T. This suggests a continuously increasing Eu moment alignment upon the increase of the magnetic field. As can be seen from the inset of Fig. 1(b), both $\mu_{sat}$ and $\mu_0 H_c$ show minima at $\phi = 45°$, consistent with that obtained above $T_N$. These observations confirm that the magnetic easy axis under the magnetic field is at $\phi = 45°$ for EuZn$_2$As$_2$ [23].

While the magnetization appears to saturate above $\mu_0 H_c$, the saturation moment $\mu_{sat}$ is lower than $\mu_{eff}$ obtained at high temperatures. To clarify this situation, we measure the magnetic torque ($\vec{\tau} = \vec{M} \times \vec{H}$) at higher fields (up to 60 T) and lower temperatures (down to $T = 0.6$ K). Figure 1(c) shows the field dependence of $\tau$ at 0.6 K at the indicated angles (the angle $\phi$ is the same as that defined in the inset of Fig. 1(a)). Two features may be seen. (1) In the low magnetic field range, there is a sharp rise to $\mu_0 H_c$, the threshold field for the saturated magnetization. (2) The torque decreases monotonically with increasing $H$ when $H > H_c$. Quantitatively, $\tau \propto H^2$ for $H > H_c$ as shown in Fig. 1(d). When all spins are polarized, torque can be expressed as $|\tau| = \chi_{eff}(\mu_0 H)^2$, where $\chi_{eff}$ is the effective volume susceptibility defined as $\chi_{eff} = \chi_{ab} - \chi_c$. Similar behavior has been observed in Fe$_3$Sn$_2$, implying that the high field $H^2$ behavior corresponds to a constant $\chi_{eff}$ [30]. Among the measured angles ($3 \leq \phi \leq 48$), both $\tau$ and $\mu_0 H_c$ decrease with increasing $\phi$, consistent with the magnetization data (Figure 1(a)). Note that the torque does not show any anomaly above $\mu_0 H_c$, suggesting neither a sudden change in the spin orientation nor any other form of metamagnetic transition.

To confirm the above observation, we perform PDO measurements of EuZn$_2$As$_2$, which is placed on a pancake coil, as shown in the inset of Fig. 1(f). This technique is sensitive to the skin or penetration depth in metals or superconductors, thus can be used to probe the electrical conductivity in such materials[31]. However, when the resistivity of the material is high, the radiofrequency field can penetrate the whole sample, and the PDO signal ($f$) becomes more sensitive to the magnetic susceptibility[32], i.e., $f \propto dM/dH$ [29]. Since EuZn$_2$As$_2$ exhibits quite high resistivity [23], the PDO signal is likely to be dominated by the magnetic properties. Figure 1(e) shows the magnetic field dependence of the PDO frequency $f$, plotted as $f - f_0$, with f$_0$ being the frequency at $H = 0$ at $T = 0.6$ K and indicated angles (the angle $\phi_{ac}$ is defined in the inset of Fig. 1(f)). The overall features are very similar to those observed in the magnetic torque, i.e., there is a sharp rise at $H_c$. However, on closer inspection, one can find new features unseen previously. Figure 1(f) presents $f(H)$ in rising (upsweep) and falling (downsweep) magnetic fields at $\phi_{ac} = 45°$. As the field rises, $f(H)$ shows staircase-like steps as indicated by arrows. The corresponding magnetic fields decrease with increasing angle $\phi_{ac}$, as shown in Figure 1(g). This and the smooth $f(H)$ measured as the field decreases suggest that the step-like behavior results from the growth of ferromagnetic domains [37,39]. The angle dependence of the critical fields corresponding to these steps is displayed in Figure 1(g). Note that these fields decrease with increasing $\phi_{ac}$, consistent with the $A$-type spin structure at zero field[33]. In other words, the FM domain alignment requires smaller fields for $H // ab$ ($\phi_{ac} = 90°$) than for $H // c$ ($\phi_{ac} = 0°$).

Given its trigonal crystal symmetry and the magnetic moment pointing to the $a$ axis in EuZn$_2$As$_2$ [23], we perform PDO measurements in another configuration: varying the field angle $\phi_{bc}$ in the $bc$ plane (30° away from the $a$ axis) as illustrated in the inset of Figure 1(h). Figure 1(h) shows the PDO frequency as a function of the magnetic field at the indicated angles ($\phi_{bc}$). While overall features are similar to those in Figure 1(e), the staircase-like behavior is absent in both the up- and down-sweeps. Instead, a hysteresis loop is observed between up-sweep and down-sweep curves. Figure 1(i) plots the difference of the frequency between up- and down-sweeps at fixed fields as a function of the angle $\phi_{bc}$. Note the magnitude of hysteresis has a similar angle dependence in all fields, peaking around $\phi_{bc} \approx 30°$. This suggests that the most difficult FM domain alignment is along $\phi_{bc} \approx 30°$, i.e., this represents the magnetic hard axis. This is further proven by plotting the angle dependence of $f$-$f_0$ at fixed fields as shown in Figure 1(j). The non-monotonic angle dependence of $f$-$f_0$ implies that the field-induced spin rearrangement depends on the direction of the applied field. The maximum |$f$-$f_0$| around 50° corresponds to large $dM/dH$.

While $f(H)$ seems at first sight smooth, without any features above $H_c$, quantum oscillations can be seen above $\mu_0 H = 20$ T after background (bg) subtraction. Figure 2(a) shows the de Haas-van Alpen (dHvA) oscillation signal $\Delta f = (f-f_0)-(f-f_0)_{bg}$ as a function of the inverse magnetic field at the indicated angles ($\phi_{ac}$). Fast Fourier transformation (FFT) reveals a single frequency $F(\phi_{ac} = 0) \approx 47$ Tesla, as illustrated by the red dashed curve in Fig. 2(b), which represents the principal oscillatory part of the Lifshitz-Kosevich (LK) formula $\Delta f \propto A\cos(F/H+\beta)$. Since $\Delta f \propto \Delta M/\Delta H$ in our case, the minimum is assigned to integer Landau Level ($N$) quantum number and the maximum to $N+1/2$. The doubling of each peak indicates Zeeman spin-splitting, which splits the Landau levels into spin-up ($N^+$) and spin-down ($N^-$) halves. A Landau fan diagram is constructed in Figure 2(c) for $\phi_{ac} = 0°$. The phase factor difference $\beta$(spin-up)-$\beta$(spin-down) = $0.31 = gm^*/2m_0$, where $g$ is the Lande factor and $m^*$ and $m_0$ are the effective and free electron masses, respectively.

To confirm the quantum oscillations observed in PDO measurements, we have further measured the magnetic field dependence of the $MR = \frac{\rho(H)-\rho(H=0)}{\rho(H=0)} \times 100$ % up to 60 T at the indicated angles [Where are these indicated?]. After careful background ($R_{bg}$) subtraction using a polynomial function, resistance oscillations may be observed above $\approx 15$ T. Figure 2(d) shows the de Shubnikov-de Hass (SdH) oscillation at T = 0.6 K and indicated angles ($\phi_{ac}$). For simplicity, we plot the oscillatory $\Delta R/R_{bg}$ as a function of $1/\mu_0 H$ at $\phi_{ac} = 5°$ and $T = 0.6$ K in Fig. 2(e). Fast Fourier transformation reveals a single frequency $F \approx 46$ Tesla, very close to that obtained from PDO data (Fig. 2(b)). The red dashed curve represents the principal oscillatory part of the LK formula $\Delta R \propto A\sin(F/H+\beta)$ where $A(H,T)$ is the oscillation amplitude and $\beta$ is the phase factor. Similar to that seen in the dHvA oscillations, the Zeeman splitting results in the doubling of peaks. A Landau fan diagram can be constructed by assigning the maximum to integer $N$ and the minimum to $N+1/2$. Figure 2(f) shows the Landau fan diagram using data displayed in Fig. 2(e) for $\phi_{ac} = 5°$ at T = 0.6 K with integer $N$ from spin-up levels ($N^+$: black dots) and spin-down levels ($N^-$: blue dots) and $N+1/2$ and $N$ levels from the LK formula simulation (red dots). We fit the data from the plot using the Lifshitz–Onsager quantization criterion $N = F/H + \beta$ [34], which gives $F = 39$ T and $\beta = 0.37$ for $\phi_{ac} = 5°$ at $T = 0.6$ K. Both $N^+(H)$ and $N^-(H)$ can be fitted by the Lifshitz-Onsager relationship with the phase factor difference $\beta$(spin-up)-$\beta$(spin-down) = $0.39 = gm^*/2m_0$.

Figure 2(g) shows the angle dependence of the averaged frequency obtained from SdH and dHvA oscillations. Compared to EuCd$_2$As$_2$, the calculated frequency $F$ for EuZn$_2$As$_2$ is larger. According to the Onsager relation $F = \left(\frac{\phi_0}{2\pi^2}\right)S$, where $\phi_0$ is the flux quantum and $S$ is the cross-sectional area of the Fermi surface normal to the magnetic field direction, a large $F$ corresponds to a large $S$. For $\phi = -1°$, the cross-section area S = 0.36 nm$^{-2}$. In a similar field configuration, S = 0.24 nm$^{-2}$ for EuCd$_2$As$_2$ [24].

The nonmonotonic angle dependence of $F$ reflects the three-dimensional (3D) shape of the Fermi surface. Our observations suggest that the Fermi surface (and therefore band structure) of EuZn$_2$As$_2$ is similar to that of EuCd$_2$As$_2$ [35]. However, the EuZn$_2$As$_2$ Eu bands are located at higher energy [32], resulting in a larger bulk band gap in EuZn$_2$As$_2$. According to Hall effect measurements, the transport in EuZn$_2$As$_2$ is dominated by holes [23]. Therefore, the observed SdH oscillations probably correspond to the hole band with an irregular ring shape also predicted for EuCd$_2$Sb$_2$ [24]. The relative sizes of the Fermi-surface cross-sections derived above implies that EuZn$_2$As$_2$ has a larger population of holes than does EuCd$_2$As$_2$. In view of the band structure of EuZn$_2$As$_2$ [35], our results therefore suggest that the Fermi energy for our sample is lower than that calculated one.

For such a hole Fermi surface, we can write β = $\Phi_B$/2π - 1/8 [34], where $\Phi_B$ is the Berry phase. The angle dependence of an average $\phi_B$ is plotted in Fig. 2(h), which shows non-zero $\Phi_B$ in all angles. This implies nontrivial electronic topology.

The temperature dependence of the SdH oscillation amplitude was measured up to 40 K and fitted to the Lifshitz-Kosevich formula Amplitude $A \propto \chi/\sinh\chi$, where $\chi = 2\pi^2 k_B T m^*/e\hbar B$, and m* = $(\hbar^2/2\pi)dS/dE$. This allows us to estimate the effective mass m* = 0.07m$_0$, which is slightly smaller than the one obtained for EuCd$_2$As$_2$. Considering the phase factor difference β(spin-up)-β(spin-down) = $gm^*/2m_0$, we can estimate g ≈ 11, which is much larger than that expected for free electrons (g = 2). Enhanced g factors have been previously observed in other Dirac and Weyl semimetals, such as ZrSiS [36], Cd$_3$As$_2$ [37], and ZrTe$_5$ [38], and attributed to strong spin-orbit coupling [38,39].

Figure 3 shows the magnetic field dependence of the $MR = \frac{\rho(H)-\rho(H=0)}{\rho(H=0)} \times 100$ % at $T = 0.6$ K up to 60 T at the indicated angles. In addition to the previously reported feature observed at low fields[23], new phenomena in the MR are observed at high fields. First, the MR at each angle remains constant between $\mu_0H_c$ and approximately 35 Tesla. Above 35 T, the MR increases gradually with increasing field, reaching ≈1000% for $H // c$. While the magnitude strongly depends on the applied field direction, with a minimum around $\phi \approx 50°$ (see the inset of Fig. 3), the MR shows no sign of saturation up to 60 Tesla. In fact, above $\mu_0H \approx 50$ T, the MR exhibits a linear field dependence. The minimum MR angle ($\phi \approx 45°$) is in accord with the torque, magnetization, and low-field MR, confirming the magnetic easy axis is close to 45°. However, considering the monotonic field dependence of the torque above $\mu_0H_c$, (Figure 1(c)) the speedy rise of the MR above ≈35 Tesla cannot not be related to any rearrangement of the magnetic moments.

At all orientations, from the Landau fan diagram shown in Fig. 2(c) and the SdH and dHvA frequencies, the quasiparticles are in the lowest Landau level by $\mu_0H \approx 50$ T. For all applied field directions, the MR increases with field without sign of saturation up to 60 T. Notably, the MR depends linearly on $H$ for $\phi = -2°$. According to Abrikosov theory [33], linear MR occurs in the quantum limit due to the linear energy-momentum dispersion relationship. Similar quantum-limit behavior has been observed in $Cd_3As_2$ above 43 T [22]. According to recent calculations [40], linear MR occurs in a fully compensated semimetal when the scattering potential is smooth within the magnetic length scale $l_m = \sqrt{(2N+1)\hbar/eH}$; effectively, the Landau wavefunctions are confined within potential minima, resulting in diffusive transport where the resistivity is proportional to $l_m^{-2} \propto H$ [40]. When $EuZn_2As_2$ reaches the 1$^{st}$ LL ($N = 1$) just below 50 T for $H // c$, $l_m \approx 6$ nm, suggesting a lengthscale of this approximate size for the disorder potential for motion within the *ab* planes. However, when $H$ is rotated towards the *ab* plane, the Landau orbits will tilt, forcing the quasiparticles to move in and out of the *ab* planes, thereby encountering a different disorder landscape. For example, for $H // ab$, the scattering potential cannot be smooth within $l_m$ due to the layered structure of $EuZn_2As_2$, which possesses an interlayer distance ≈ 0.7175 nm [23]. It is the variation of the typical lengthscale of the potential landscape encountered by the tilted cyclotron orbits that results in the observed angle dependence of the MR.

In summary, we have experimentally investigated the electronic and magnetic properties of $EuZn_2As_2$, a magnetic topological semimetal candidate, in pulsed magnetic fields of up to 60 T. Giant positive magnetoresistance (MR) is observed above $\mu_0H \approx 20$ T; however, magnetic torque and proximity detector oscillator (PDO) measurements show no corresponding anomaly. By analyzing the Shubnikov-de Haas and de Haas-van Alphen oscillations between 15 and 50 T, we obtain the oscillation frequency $F$ = 46 Tesla for $H \mathbin{/\mkern-3mu/} c$, whilst observing clear Zeeman splitting of the oscillations attributable to a substantially enhanced $g$-factor. The oscillation frequency extracted from both SdH and dHvA oscillations varies non-monotonically with the angle between the $c$ axis and the $ab$ plane, demonstrating the complex hole band shape of $EuZn_2As_2$. While the Berry phase extracted from the Landau phase diagram varies with angle as well, a consistently non-zero $\Phi_B$ indicates the non-trivial topology of this hole band.

From both the quantum-oscillation frequencies and Landau fan diagrams, it is clear that only the first Landau level is occupied by 50 T. Above this field, the MR varies linearly with $H$, without any sign of saturation. This is attributable to the quantum-limit mechanism originally due to Abrikosov and further developed in [40], where compressed Landau wavefunctions are confined by smooth potential fluctuations, resulting in diffusive transport. The field-angle dependence of the large MR demonstrates the varying disorder potential lengthscales that exist in layered $EuZn_2As_2$. Our findings help understand the interplay between magnetism, topological properties, disorder, and emergent behavior in the quantum limit.

**Acknowledgments:**

Work at University of South Carolina and Louisiana State University was supported by NSF through Grant DMR-1504226. A portion of this work was performed at the National High Magnetic Field Laboratory (NHMFL), which is supported by National Science Foundation Cooperative Agreement Nos. DMR- 1644779 and the Department of Energy (DOE). J.S. acknowledges support from the DOE BES program "Science at 100 T," which permitted the design and construction of specialized equipment used in the high-field studies.

**Author contributions**

R. J. and J. B. proposed and designed the research. J. B. and W. X. synthesized the single crystals. J.B. carried out the low magnetic field measurements and data analysis. J. S., J. B., and S. S. carried out the high magnetic field measurements and data analysis. J. B., R. J., and J. S. wrote the paper with input from all co-authors. R.J. oversaw the project.

**Data availability:**

The source data and related supporting information are available upon reasonable request from the corresponding author.

**Competing interests:**

The authors declare no competing interests.

**Correspondence** and requests for materials should be addressed to Rongying Jin

**Fig. 1. Magnetic properties** (a) Inverse magnetic susceptibility versus temperature at various angles. Inset: The Curie-Weiss temperature and effective magnetic moment versus angle. (b) Magnetization as a function of magnetic field at ϕ = 45°. Inset: the saturation moment and magnetic field transition versus angle. (c) Magnetic field dependence of torque taken at various angles. (d) torque versus $H^2$ with a dashed line for eye guidelines. (e) PDO frequency as a function of magnetic field taken at various angles. (f) Low magnetic field region of PDO frequency at ϕ = 45° pointing the step-like behavior. Inset: PDO pancake coil (g) The magnetic field value at each step versus magnetic field. (h) PDO signal as a function of magnetic field taken at various angles. (i) normalized difference between raising and falling field curves as a function of angle taken at fixed magnetic field. (j) PDO frequency versus angle at fixed magnetic field.

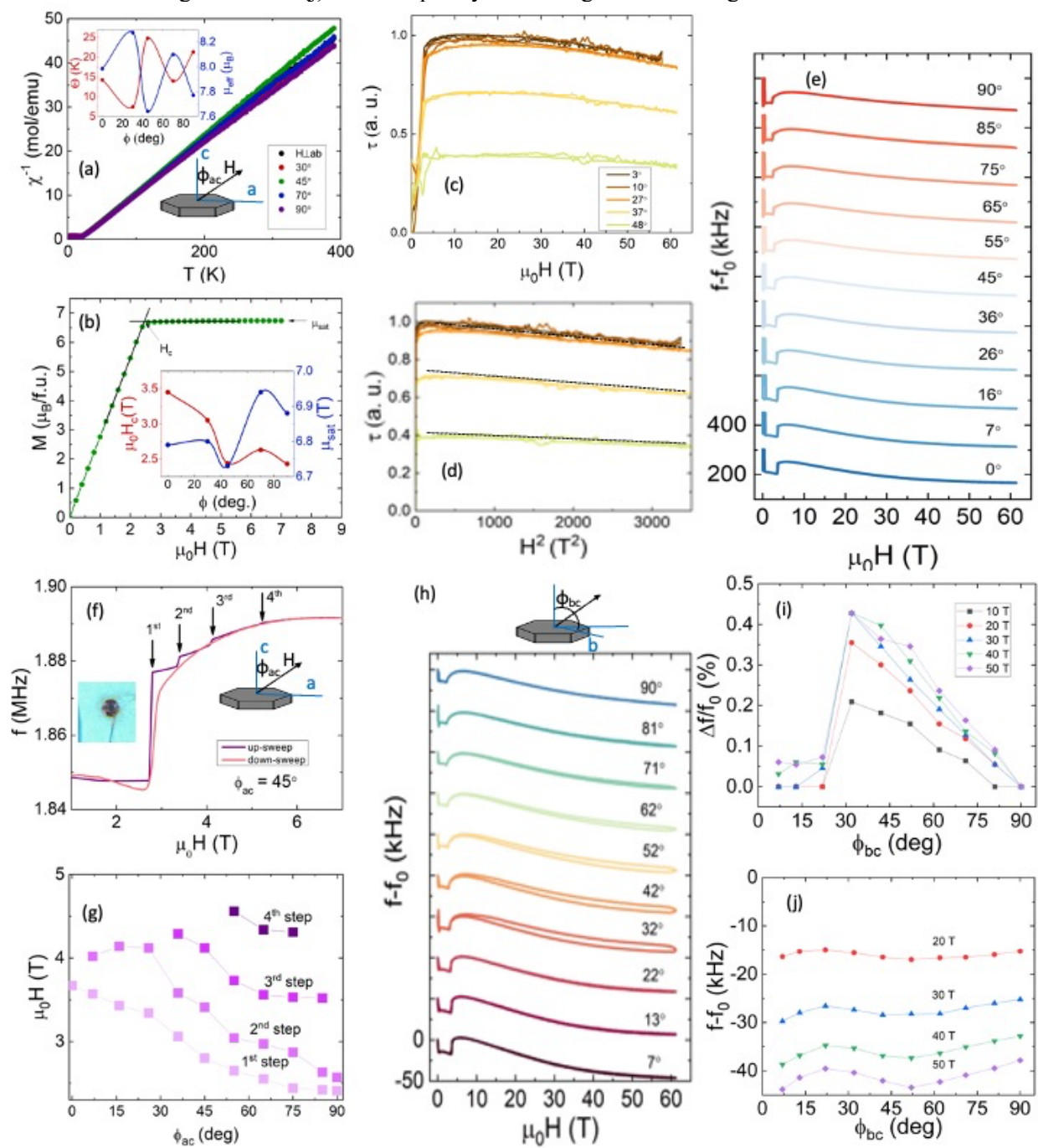

**Fig. 2. Quantum oscillations** (a) SdH quantum oscillation as a function of magnetic field at various angles. (b) dHvA quantum oscillation as a function of magnetic field at various angles. (c) SdH quantum oscillation as a function of inverse magnetic field at $\phi = 5°$ (blue line) and the simulation of SdH oscillation described by LK formula (red dashed line) (d) Landau fan diagram for oscillation at $\phi = 5°$. (e) dHvA quantum oscillation as a function of inverse magnetic field at $\phi = 0°$ (blue line) and the simulation of SdH oscillation described by LK formula (red dashed line). (f) Landau fan diagram for oscillation at $\phi = 0°$. (g) Frequency versus angle (h) Berry phase as a function of angle.

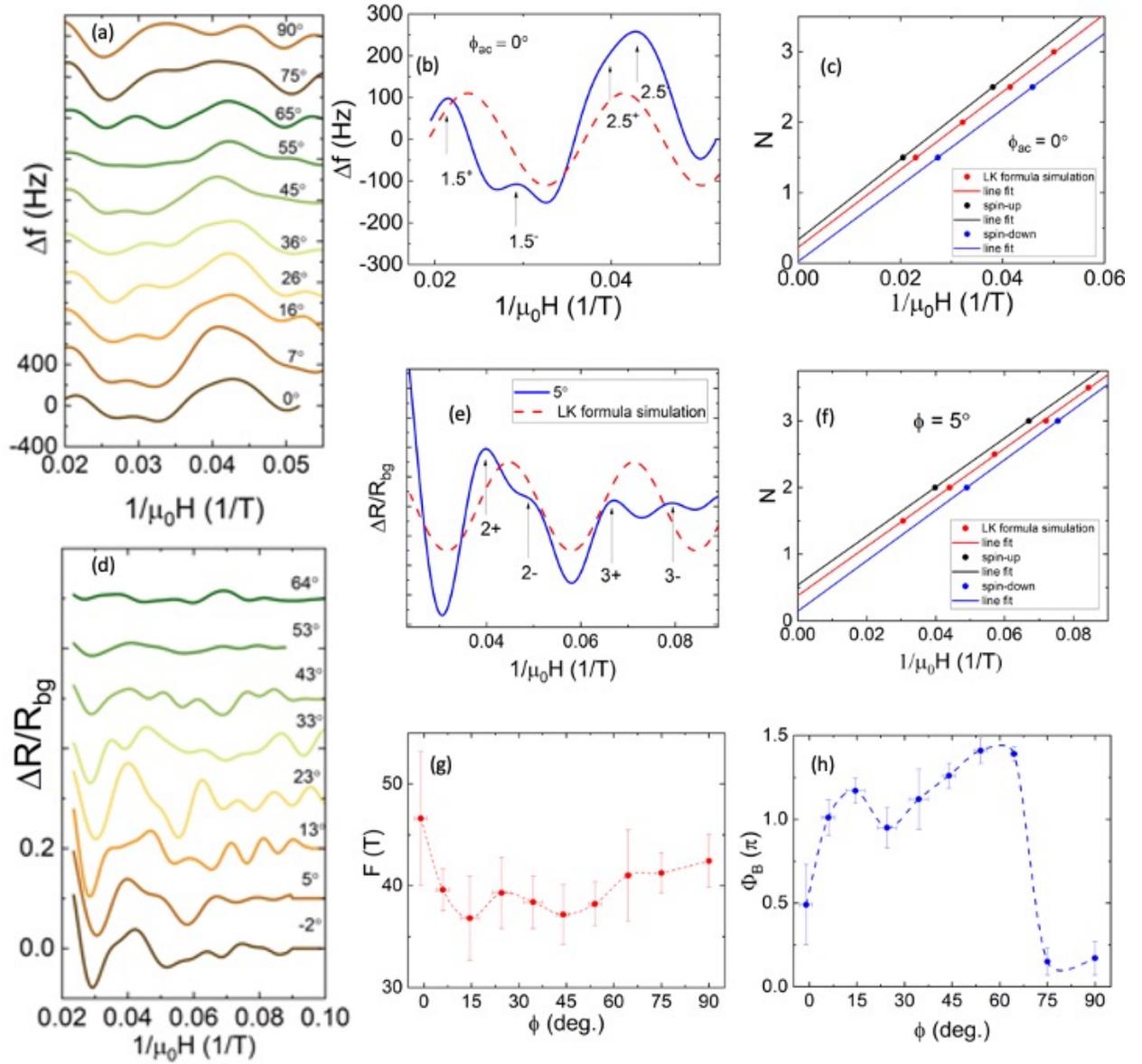

**Fig. 3. Linear Quantum MR.** Magnetoresistivity as a function of the magnetic field at various angles. Inset: MR at fixed magnetic field versus angle at T = 0.6 K.

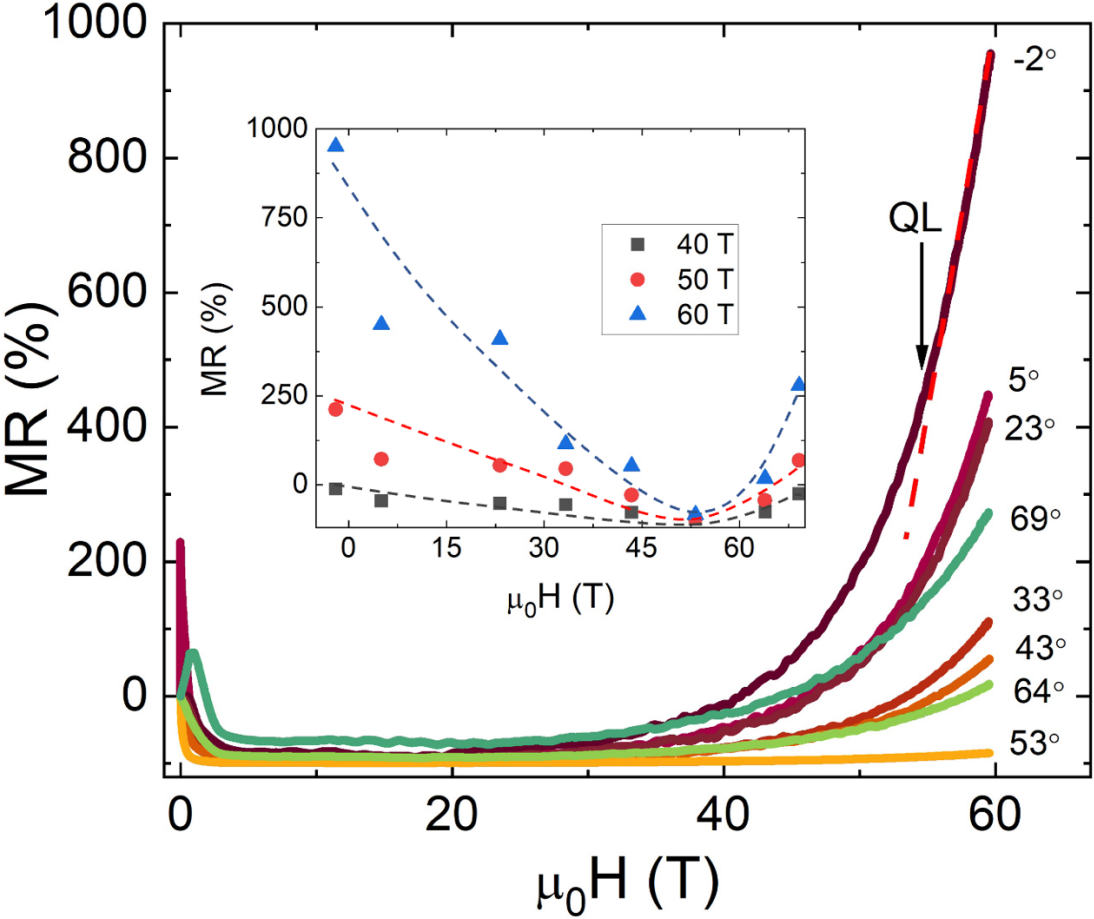